# Interactive zoom display in smartphone-based digital holographic microscope for 3D imaging

Yuki Nagahama[1,*]

Institute of Engineering, Tokyo University of Agriculture and Technology, 2-24-16 Naka-cho, Koganei, Tokyo 184-8588, Japan
[*] yuki-nagahama@go.tuat.ac.jp

## Abstract

Digital holography has applications in bio-imaging because it can simultaneously obtain the amplitude and phase information of a microscopic sample in a single shot, thus facilitating non-contact, noninvasive observation of the 3D shape of transparent objects (phase objects, which can be mapped with the phase information,) and moving objects. The combination of digital holography and microscopy is called digital holographic microscopy (DHM). In this study, we propose a smartphone-based DHM system for 3D imaging that is compact, inexpensive, and capable of observing objects in real time; this system includes an optical system comprising a 3D printer using commercially available image sensors and semiconductor lasers; further, an Android-based application is used to reconstruct the holograms acquired by this optical system, thus outlining the amplitude and phase information of the observed object. Also, by utilizing scalable diffraction calculation methods and touchscreen interaction, we implemented zoom functionality through pinch-in gestures. The study results showed that the DHM system successfully obtained the amplitude and phase information of the observed object via the acquired holograms in an almost real time manner. Thus, this study showed that it is possible to construct a low cost and compact DHM system that includes a 3D printer to construct the optical system and a smartphone application to reconstruct the holograms. Furthermore, this smartphone-based DHM system's ability to capture, reconstruct, and display holograms in real time demonstrates its superiority and novelty over existing systems. This system is also expected to contribute to biology fieldwork and pathological diagnosis in remote areas.

## Introduction

Digital holography [1] is a method of recording light waves emitted from an object as holograms and reconstructing the holograms [2] using light wave propagation calculations to observe the object in three dimensions. It has applications in bio-imaging and other fields [3] because it can simultaneously obtain the amplitude and phase information of a sample in a single shot, facilitating non-contact, noninvasive observation of the 3D shape of transparent objects (phase objects) and moving objects. The combination of digital holography and microscopy is called digital holographic microscopy (DHM).

Research on DHM as a tool for bio-imaging is being conducted at various research institutions; such research efforts regarding DHM include observing cell dynamics such as cell division, membrane dynamics of red blood cells, stem cell dynamics, and diagnosing sickle cell disease [4-7]. To date, many DHM systems have been custom built for specific applications. However, most of these systems are large and expensive. In contrast, a low cost portable DHM system has many applications, including biological fieldwork and pathological diagnosis in remote areas. Therefore, miniaturization of the optical system of DHM has been proposed by various groups of researchers [8-12]. Additionally, a miniaturized system that includes replacing the image sensor and computational image reconstruction system in the optical system of DHM with a smartphone has been proposed [13-15]. However, because the shape and other image sensor parameters (pixel pitch and number of pixels) differ depending on the smartphone used, it is necessary to design a component for establishing a connection between the optical system and the smartphone and a component related to the image sensor of the computation system for each smartphone. Also, DHM in which a smartphone and optical system are connected via USB has also been proposed [16].

However, existing smartphone-based DHMs only capture holograms on the smartphone and reconstruct the holograms on another device [13, 14, 16] or capture holograms and then reconstruct them non-real-time on the smartphone [15], and no system has been proposed that captures and reconstructs holograms in real time on a smartphone.

Hence, in this study, we aim to build a DHM system in which a reconstructed image of hologram generated on the image sensor of a USB camera built into an optical system is observed by an Android smartphone that serves as the computational image reconstruction system in real time. Furthermore, the reconstructed hologram is displayed and zoomed interactively through pinch-in gestures. Additionally, we assess whether the designed system can acquire and reconstruct holograms successfully. Furthermore, we verify the frame rate at which the constructed system operates.

## Method

We constructed a Gabor-type optical system as the optical system the proposed DHM (Figure 1). The optical system was manufactured using a 3D printer; the printer's dimensions, excluding cables, were approximately 101 [mm] (width) × 50 [mm] (depth) × 55 [mm] (height). The image sensor used was a disassembled commercially available USB camera (ELECOM UCAM-C980FBBK).

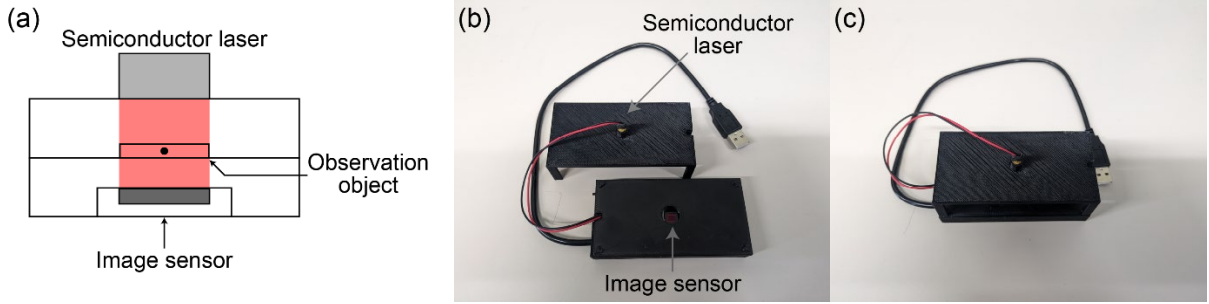

Figure 1. (a) Schematic of the Gabor-type optical system. (b) The optical system of the proposed DHM. (c) The optical system of the proposed DHM in operation.

Smartphones have limitations in terms of computational performance and memory capacity. Therefore, we adopted band-limited double-step Fresnel diffraction (BL-DSF) [17], which can reduce the number of data points, thus facilitating faster computational image reconstruction from holograms. In Fourier optics, diffraction calculations can be classified into two types. The first is convolution-based diffraction and the second is Fourier transform-based diffraction. Here, we present the angular spectrum method (ASM) as an example of convolution-based diffraction. The ASM is shown below:

$$u_2(x_2, y_2) = \text{FFT}^{-1}\left[\text{FFT}[u_1(x_1, y_1)]\right]\exp\left(-2\pi i z \sqrt{1/\lambda^2 - f_x^2 - f_y^2}\right) \quad (1)$$

Here, $\lambda$ is the wavelength of light, and $\text{FFT}[\cdot]$ and $\text{FFT}^{-1}[\cdot]$ are fast Fourier transform and inverse fast Fourier transform operators, respectively. Further, $u_1(x_1, y_1)$ and $u_2(x_2, y_2)$ indicate the source and destination planes, respectively. Additionally, $(f_x, f_y)$ are coordinates in the frequency domain and $z$ is the propagation distance. The advantage of convolution-based diffraction is that the source and destination planes have the same sampling rate. However, because FFT convolution is a circular convolution, to use it as a linear convolution, the following is necessary: to extend the source and destination planes with zero padding and expand the size to $2N \times 2N$ ($N$ is the number of pixels in the horizontal and vertical directions of the hologram). Therefore, the memory usage and computational cost of the ASM are proportional to $4N^2$ and $4N^2 \log 4N$, respectively. Thus, this method uses more memory and has a longer computational period.

To overcome this problem, double-step Fresnel diffraction (DSF) has been proposed [18]. This diffraction method includes computing the propagation of light from the source plane to the destination plane through a virtual plane $(x_v, y_v)$ by two Fourier transform-based diffraction calculations. Because DSF is based on Fourier transform-based diffraction, it does not require zero padding. Additionally, most Fourier transform-based diffraction methods change the sampling rate of the source and destination planes. However, in DSF, by adjusting the distance $z_1$ from the source surface to the virtual plane and the distance $z_2$ from the virtual plane to the destination surface, the sampling rate of the source surface and the destination surface can be arbitrarily determined. If the sampling rate of the source plane is $p_s$ and the sampling rate of the destination plane is $p_d$, the relationship between $p_s$ and $p_d$ is $p_d = |z_1/z_2|p_s$. Furthermore, BL-DSF introduces a rectangular function for band-limiting to avoid aliasing that occurs under certain conditions. BL-DSF is expressed as follows:

$$u_2(m_2, n_2) = C_{z_2} \text{FFT}^{sgn(z_2)} \left[ \exp\left(\frac{i\pi z(x_v^2 + y_v^2)}{\lambda z_1 z_2}\right) \text{Rect}\left(\frac{x_v}{x_v^{\max}}, \frac{y_v}{y_v^{\max}}\right) \text{FFT}^{sgn(z_1)} \left[ u_1(m_1, n_1) \exp\left(\frac{i\pi(x_1^2 + y_1^2)}{\lambda z_1}\right) \right] \right] \quad (2)$$

Here, $z_1$ is the propagation distance from the source plane to the virtual plane, and $z_2$ is the propagation distance from the virtual plane to the destination plane; $C_{z_2} = \exp\left(\frac{i\pi}{\lambda z_2}(x_2^2 + y_2^2)\right)$. The operator $\text{FFT}^{sgn(z)}$ indicates a forward FFT if the sign of $z$ is positive and an inverse FFT if the sign of $z$ is negative. A previous study outlines how to determine the bandwidth restriction area [17]. The format of BL-DSF is not convolutional. Therefore, the memory usage and computational cost of BL-DSF are proportional to $N^2$ and $N^2 \log N$, respectively.

The Android OS itself provides a function to acquire images from a USB camera on an Android smartphone [19]; however, whether this function is implemented varies depending on the device. Therefore, in this study, we used a library called UVCCamera [20] to acquire images from a USB camera on an Android smartphone. Additionally, the component that processes holograms taken with a USB camera is implemented using OpenCV [21], and the component that performs computational image reconstruction from holograms, which requires a particularly high computation speed, is implemented using C++ with Android NDK [22] instead of Java. Figure 2 shows how programming languages and libraries are used. For convenience, the Android application developed this time is called PocketHoloScope.

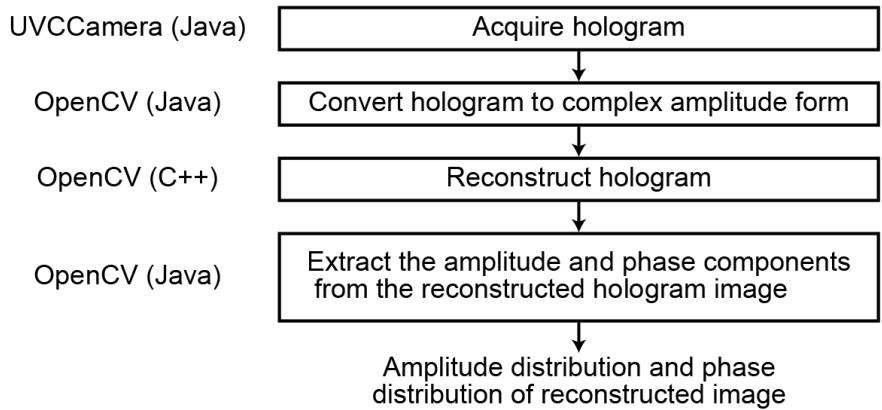

Figure 2. Flow from hologram acquisition to reconstruction and the associated libraries and languages used in PocketHoloScope.

## Results

Figure 3 shows a screenshot of PocketHoloScope. Table 1 shows the conditions for the hologram reconstruction computation. The resolution of the image sensor of the USB camera is 3,264 × 2,448 pixels; however, this resolution is significantly large to acquire, reconstruct, and display the hologram. Therefore, down sampling is performed at the stage of acquiring the hologram, and the resolution of the hologram is set to 1920 × 1,440 pixels. Note that Google Pixel 7a (Android version 14) was used as the execution environment for PocketHoloScope.

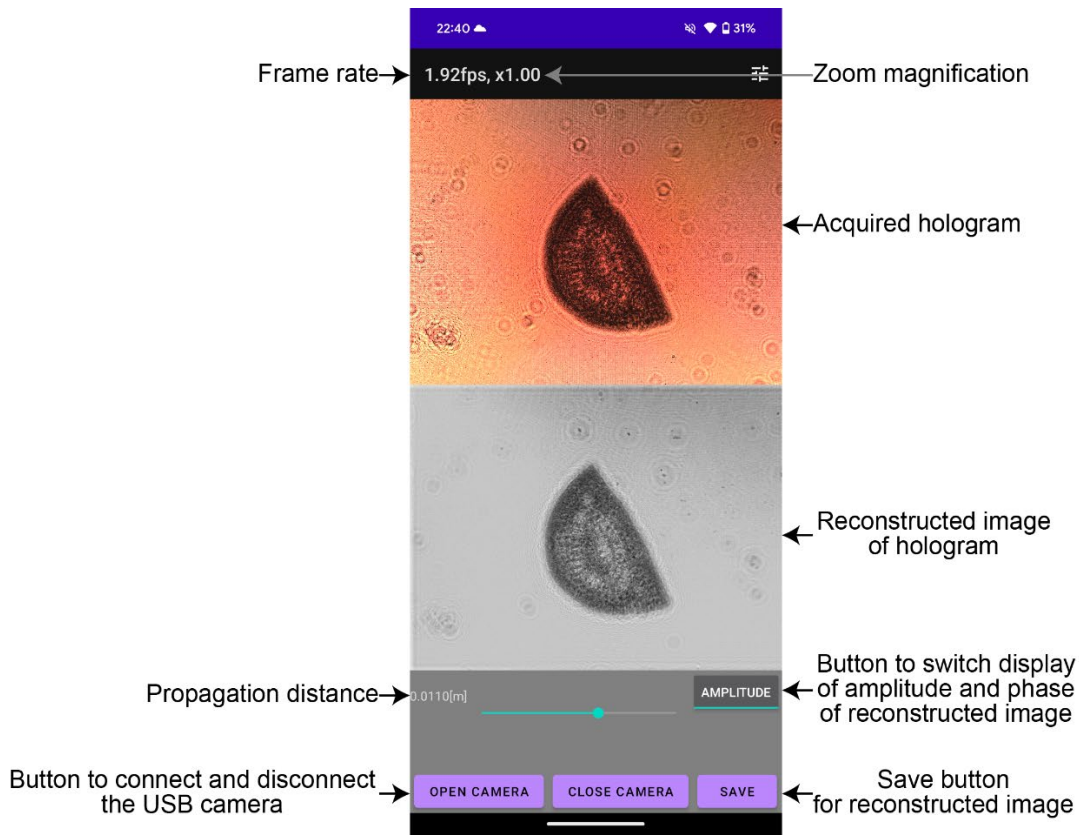

Figure 3. Screenshot of PocketHoloScope

Table 1. Hologram reconstruction computation conditions

| Image sensor resolution | 3,264 × 2,448 pixels |
|---|---|
| Image sensor pixel pitch | 1.47 μm |
| Hologram resolution | 1920 × 1,440 pixels |
| Hologram sampling rate | 2.50 μm |
| Laser wavelength | 650 nm |

Figure 4(a) shows the amplitude component of the reconstructed image using the BL-DSF when the observed object is pine leaf, c.s., and the light propagation distance is focused on the observed object (0.011 m). Additionally, the measured frame rate was observed to be 1.92 fps. For reference, Figure 4(b) shows the amplitude component of the reconstructed image of a hologram that was reconstructed using the ASM. The frame rate when reconstructing the hologram using the ASM was 0.72 fps. In addition, Figure 4(c) shows the phase component of the reconstructed image using the BL-DSF and Figure 4(d) shows the phase component of the reconstructed image using the ASM.

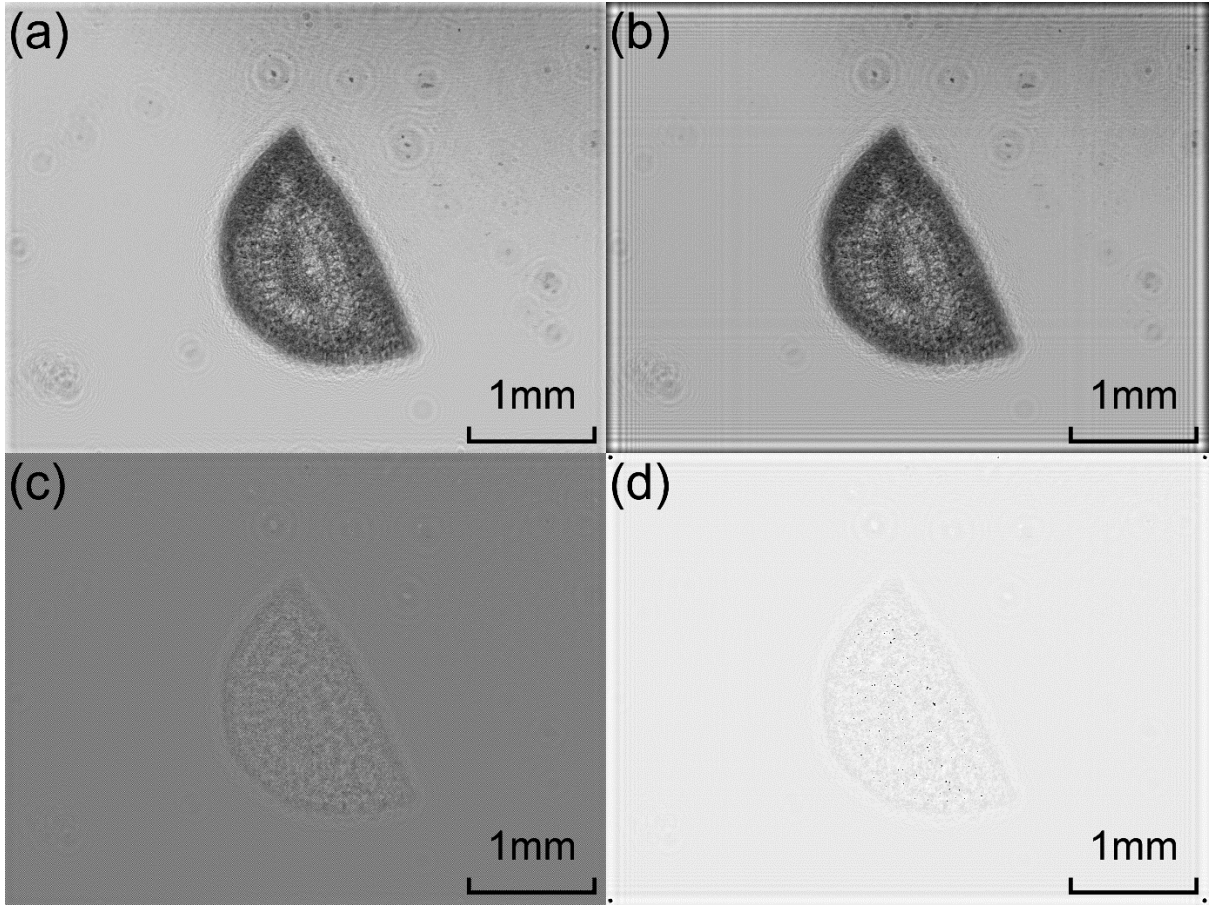

Figure 4. (a) Amplitude component of reconstructed hologram image of pine leaf, c.s. (BL-DSF). (b) Amplitude component of reconstructed hologram image of pine leaf, c.s. (ASM). (c) Phase component of reconstructed hologram image of pine leaf, c.s. (BL-DSF). (d) Phase component of reconstructed hologram image of pine leaf, c.s. (ASM).

BL-DSF allows the sampling rate to be changed before and after diffraction calculations and can be used to enlarge or reduce the reconstructed image of the hologram. Figures 5(a) and (b) show the amplitude and phase components of the reconstructed image enlarged 1.2 times from the hologram. Furthermore, Figures 5(c) and (d) show the amplitude and phase components of the reconstructed image reduced 0.8 times from the hologram.

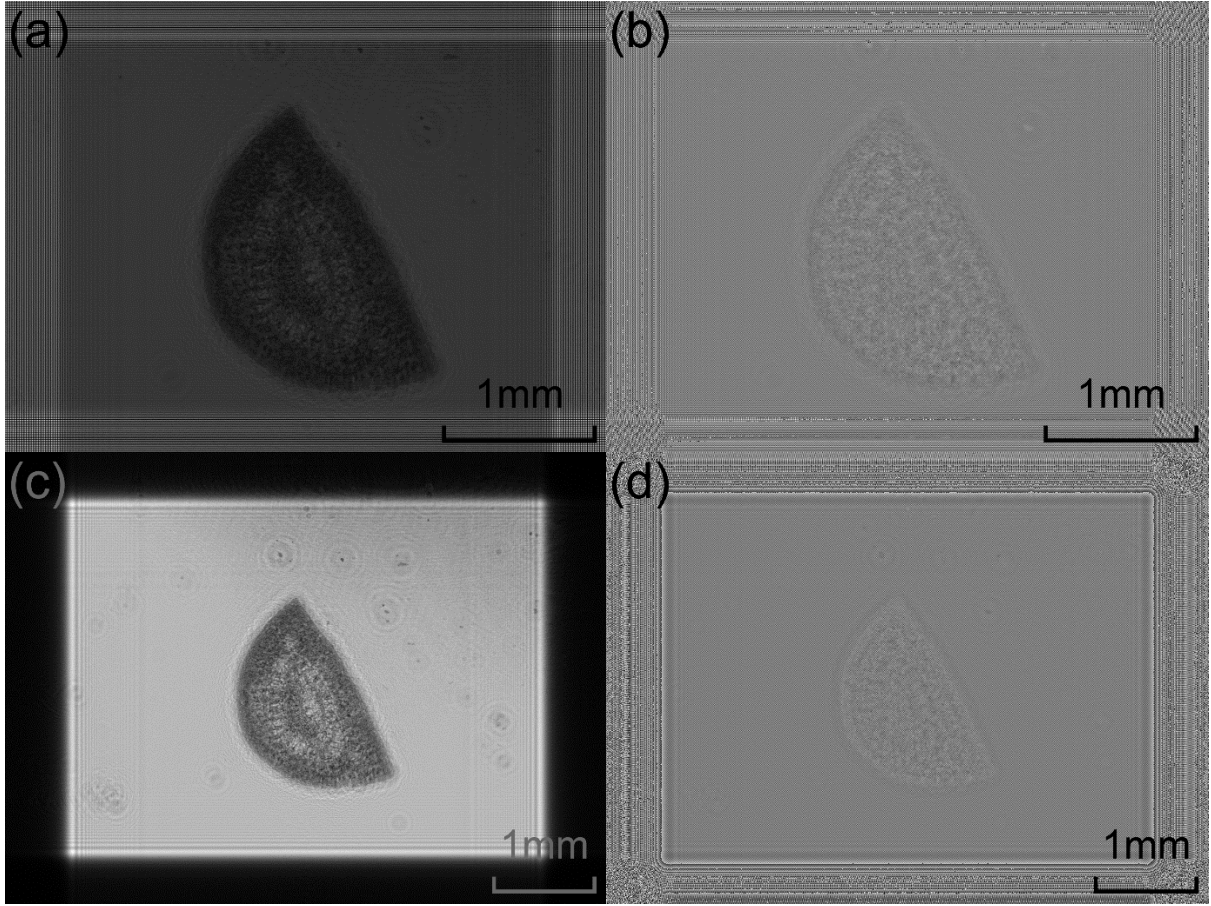

Figure 5. (a) Amplitude component of the reconstructed image enlarged 1.2 times from the hologram. (b) Phase component of the reconstructed image enlarged 1.2 times from the hologram. (c) Amplitude component of the reconstructed image reduced 0.8 times from the hologram. (d) Phase component of the reconstructed image reduced 0.8 times from the hologram.

We have implemented a function to zoom the reconstructed hologram image using a pinch-in gesture on the touch screen of a smartphone. The results are shown in Appendix 1.

## Discussion

We discuss the frame rate of the proposed system. As mentioned in the "Results" section, the DHM system proposed in this study acquired, reconstructed, and displayed holograms at a frame rate of 1.92 fps when using BL-DSF and at that of 0.72 fps when using ASM. Considering that the computation cost of BL-DSF is 1/4 that of ASM, this result can be considered to be acceptable considering the difference in frame rate and other calculations. Additionally, when employing BL-DSF with the frame rate of 1.92 fps, the observed images were displayed in almost real time when observing stationary objects.

In this research, we implemented and demonstrated a function that uses the pinch-in gesture on the touch panel of a smartphone to zoom in on the reconstructed hologram image. This result can be said to demonstrate superiority and novelty compared to the smartphone-based DHM in previous research in that it enabled interactive operation by capturing, playing, and displaying holograms in real time.

In the future, to observe moving objects in real-time, enhancing the system's frame rate will be essential. To achieve this, it is considered necessary to implement parallel processing using multiple CPU threads and GPUs.

## Data availability

The data and codes supporting the findings of this study are available from the corresponding author upon reasonable request.

## Figure Legends

Figure 1. (a) Schematic of the Gabor-type optical system. (b) The optical system of the proposed DHM. (c) The optical system of the proposed DHM in operation.

Figure 2. Flow from hologram acquisition to reconstruction and the associated libraries and languages used in PocketHoloScope

Figure 3. Screenshot of the developed Android application

Figure 4. (a) Amplitude component of reconstructed hologram image of pine leaf, c.s. (BL-DSF). (b) Amplitude component of reconstructed hologram image of pine leaf, c.s. (ASM). (c) Phase component of reconstructed hologram image of pine leaf, c.s. (BL-DSF). (d) Phase component of reconstructed hologram image of pine leaf, c.s. (ASM).

Figure 5. (a) Amplitude component of the reconstructed image enlarged 1.2 times from the hologram. (b) Phase component of the reconstructed image enlarged 1.2 times from the hologram. (c) Amplitude component of the reconstructed image reduced 0.8 times from the hologram. (d) Phase component of the reconstructed image reduced 0.8 times from the hologram.

## Competing interests

The author declares no competing interests.

## Acknowledgments

This work is supported by the Japan Society for the Promotion of Science (JSPS) KAKENHI (Grant-in-Aid for Early-Career Scientists 21K17760) 2021.The authors would like to thank Maruzen-Yushodo Co., Ltd. (https://kw.maruzen.co.jp/kousei-honyaku/) for the English language editing.

## Author contributions

Y.N. wrote the main manuscript text, prepared Figs. 1–5, performed the experiments, and reviewed the manuscript.